\documentclass[fleqn,usenatbib]{mnras}

\usepackage{mathptmx}

\usepackage[T1]{fontenc}
\usepackage{ae,aecompl}

\usepackage{amsmath}	
\usepackage{graphicx}	

\usepackage{fontawesome}
\usepackage{newtxtext,newtxmath}

\newcommand{\numobserved}{57}
\newcommand{\numconfirmed}{15}

\newcommand{\nummodeled}{20}
\newcommand{\numyears}{three}

\newcommand{\editref}[1]{{{#1}}}
\newcommand{\editreft}[1]{{#1}}
\newcommand{\editrefs}[1]{{#1}}

\newcommand{\editaa}[1]{#1}
\newcommand{\editas}[1]{#1}
\newcommand{\editast}[1]{#1}

\title[Doubly imaged quasars]{\editaa{High-resolution imaging follow-up of doubly imaged quasars}}

\author[Shajib et al.]{
	Anowar J. Shajib$^{1, 2}$\thanks{ajshajib@uchicago.edu}, 
    Eden Molina$^{1}$,  
    Adriano Agnello$^{3}$,  
    Peter R. Williams$^{1}$, \and 
    Simon Birrer$^{1,4}$,
    Tommaso Treu$^{1}$\thanks{Packard fellow},
    Christopher D. Fassnacht$^{5, 6}$,
    Takahiro Morishita$^{7}$, \and
    Louis Abramson$^{8}$, 
    Paul L. Schechter$^{9}$,
    Lutz Wisotzki$^{10}$
    \\ \\
$^{1}$Department of Physics and Astronomy, University of California, Los Angeles, CA 90095, USA \\
$^{2}$Department  of  Astronomy  \&  Astrophysics,  University  of Chicago, Chicago, IL 606374, USA \\
$^{3}$DARK, Niels Bohr Institute, University of Copenhagen, Jagtvej 128, 2200 Copenhagen, Denmark \\
$^{4}$Kavli Institute for Particle Astrophysics and Cosmology and Department of Physics, Stanford University, Stanford, CA 94305, USA \\
$^{5}$Physics Department, UC Davis, 1 Shields Ave., Davis, CA 95616, USA \\
$^{6}$Carnegie Visiting Scientist \\
$^{7}$Space Telescope Science Institute, 3700 San Martin Drive, Baltimore, MD 21218, USA \\
$^{8}$Carnegie Observatories, 813 Santa Barbara Street, Pasadena, CA 91101, USA \\
$^{9}$MIT Kavli Institute for Astrophysics and Space Research, Cambridge, MA 02139, USA \\
$^{10}$Leibniz-Institut f\"{u}r Astrophysik Potsdam (AIP), An der Sternwarte 16, 14482 Potsdam, Germany
}

\date{Accepted XXX. Received YYY; in original form ZZZ}

\pubyear{2020}

\begin{document}

\label{firstpage}
\pagerange{\pageref{firstpage}--\pageref{lastpage}}
\maketitle

\begin{abstract}

We report upon \numyears\ years of follow-up and confirmation of doubly imaged quasar lenses through imaging campaigns from 2016--2018 with the Near-Infrared Camera2 (NIRC2) on the W. M. Keck Observatory. 
A sample of \numobserved\ quasar lens candidates {are} imaged in adaptive-optics-assisted or seeing-limited \editast{$K^\prime$-}band observations. Out of these \numobserved\ candidates, \numconfirmed\ {are} confirmed as lenses. \editas{We form a sample of \nummodeled\ lenses adding in a number of previously-known lenses that were imaged with NIRC2 in 2013--14 as part of a pilot study. By modelling these 20 lenses, we} obtain \editast{$K^\prime$-}band relative photometry and astrometry of the quasar images and the lens galaxy. We also provide the lens properties and predicted time delays to aid planning of follow-up observations necessary for various astrophysical applications, \editas{e.g., spectroscopic follow-up to obtain the deflector redshifts for the newly confirmed systems.}
\editref{We compare the departure of the observed flux ratios from the smooth-model predictions between doubly and quadruply imaged quasar systems. We find that the departure is consistent between these two types of lenses if the modelling uncertainty is comparable.}

\end{abstract}


\begingroup
\let\clearpage\relax
\endgroup
\newpage

\section{Introduction}
\label{intro}

Strong gravitational lensing is 
\editaa{the production of multiple images of a distant object} due to gravitational \editast{deflection of light by} a foreground massive object. When this foreground massive object is a galaxy, we refer to these systems as galaxy-scale lenses (hereafter, lenses). Lenses are useful for their numerous astrophysical applications -- from quantifying the dark matter and baryonic fraction in galaxies to resolved studies of distant lensed sources \citep[e.g.,][]{Falco99, Auger10, Sonnenfeld15, Sonnenfeld18b}. Strongly lensed quasars are particularly useful for measuring the Hubble constant, detecting dark matter substructure, \editas{and} studying the stellar initial mass function \citep[e.g.,][]{Nierenberg14, Schechter14, Shajib20, Birrer20}.


Despite the usefulness of lensed quasars, these systems are relatively rare, with $\sim$200 discovered so far \editaa{making up a very heterogeneous sample} \citep{Lemon19} \editaa{and much brighter than what in principle is allowed by the depth of current imaging surveys \citep{Treu18}}. \editaa{\editas{Moreover}, each science case\editas{ -- }e.g.\editas{,} galaxy masses, cosmography\editas{ -- }has rather stringent requirements on the lensing configuration and quality of ancillary data. Therefore, assembling large and \textit{complete} samples of lensed quasars is still an active effort.} \editas{To expedite the discovery of these rare systems}, multiple techniques of data-mining from large-area sky surveys -- such as the Sloan Digital Sky Survey \editast{\citep[SDSS;][]{York00}}, the VLT Survey Telescope ATLAS \editast{\citep[VST-ATLAS;][]{Shanks15}}, the Dark Energy Survey \editast{\citep[DES;][]{Flaugher15}} -- have been recently developed \citep[e.g.,][]{Agnello15, Williams17, Williams18, Agnello18b}. Most recently, the combination of ESA-\textit{Gaia}'s high spatial resolution and population-mixture selection techniques on ground-based survey data has led to new lensed quasar discoveries in the DES, the Panoramic Survey Telescope and Rapid Response System \editast{\citep[Pan-STARRS;][]{Chambers16}}, and the Kilo-Degree Survey \editast{\citep[KiDS;][]{deJong13}} footprints \editas{\citep[e.g.,][]{Agnello18c, Spiniello18, Treu18, Lemon19}}.


To extract scientific information from these strongly lensed quasar systems, dedicated follow-up observations are necessary. First, spectroscopic \editast{observations are} required to obtain the redshifts of the deflector and the source. Second, multi-band high-resolution imaging is necessary to obtain robust photometry and astrometry, and to model the mass distribution in the deflector galaxy \citep[e.g.,][]{Shajib19}. Third, long-term monitoring is needed to measure the time-delays between the quasar images for cosmographic applications \citep[e.g.,][]{Eigenbrod05}.


In this paper, we report \editaa{on} a sample of confirmed doubly imaged quasar lenses (hereafter, doubles) from a follow-up imaging campaign \editast{obtained over a \numyears\ year period}. We acquired \numobserved\ lens candidates from data-mining through various surveys. We followed up these candidates with the Near-Infrared Camera2 (NIRC2) imager on the W. M. Keck Observatory. 
These observations enabled us to identify lensed arcs and rings in part of the sample, as well as the confirmation of small-separation lenses (down to $\sim$0.3--0.5 arcsecond). Out of these \numobserved\ candidates, we {confirm} \numconfirmed\ as doubles. We model these doubles and provide astrometry, photometry, and inferred lens properties to facilitate future planning of follow-ups to obtain ancillary data. \editaa{\editas{We also present data and models from} a previous pilot program in 2013--14, where 7 known lensed quasars were imaged with NIRC2 in search of \editas{lensed arcs} from the quasar host galaxies.}


This paper is organized as follows. In Section \ref{sec:nirc2_campaign}, we describe the imaging campaign to follow up and confirm lensed quasar candidates. Then in Section \ref{sec:modelling}, we explain the modelling procedure for the sample of confirmed doubles. In Section \ref{sec:results}, we provide astrometry, photometry, and lens properties for the sample of doubles. Finally in Section \ref{sec:discussion}, we conclude the paper.

%
%

\section{Keck NIRC2 imaging campaigns} \label{sec:nirc2_campaign}

\renewcommand{\arraystretch}{1.2}
\begin{table*}
\caption{\label{tab:lens_obs}
	Coordinates, expossure time, and observation date for the modeled sample of doubly imaged quasars.
}
\begin{tabular}{lrrlll}
\hline
Name & RA & dec & Exposure time & Observation date & Reference(s) \\
& (deg) & (deg) & (seconds) & \\
\hline
HE 0013$-$2542 & 3.93292 & -25.43806 & 300 & 2016 Sept 21 & \editref{This paper} \\
HE 0047$-$1756 & 12.61580 & -17.66930 & 3780 & 2013 Aug 31 & \citet{Wisotzki04} \\
PS J0140+4107 & 25.20420 & 41.13331 & 1500 & 2018 Jan 02 & \citet{Lemon18} \\
Q0142-100 & 26.31940 & -9.75475 & 1020 & 2013 Aug 30 & \citet{Surdej87} \\
WGA 0235$-$2433 & 38.86426 & -24.55368 & 420 & 2017 Oct 04 & \citet{Agnello18b, Lemon18} \\
WGD 0245$-$0556 & 41.35651 & -5.95015 & 1740 & 2018 Jan 02 & \citet{Agnello18c} \\
SDSS J0246$-$0825 & 41.55083 & -18.75139 & 2160 & 2013 Aug 30 & \citet{Inada05} \\
PS J0417+3325 & 64.49683 & 33.41700 & 1140 & 2018 Jan 02 & \citet{Lemon18} \\
SDSS J0806+2006 & 121.59867 & 20.10874 & 1500 & 2014 Mar 18 & \citet{Inada06} \\
PS J0840+3550 & 130.13842 & 35.83334 & 1140 & 2018 Jan 02 & \citet{Lemon18} \\
PS J0949+4208 & 147.47830 & 42.13381 & 1140 & 2018 Jan 02 & \citet{Lemon18} \\
SDSS J1001+5027 & 150.36876 & 50.46595 & 3240 & 2014 Mar 19 & \citet{Oguri05} \\
LBQS 1009$-$0252 & 153.06625 & -3.11750 & 420 & 2018 Jan 02 & \citet{Hewett94} \\
SDSS J1128+2402 & 172.07705 & 24.03817 & 1140 & 2018 Jan 02 & \citet{Inada14} \\
SDSS J1650+4251 & 252.68100 & 42.86369 & 1620 & 2013 Aug 30 & \citet{Morgan03} \\
HS 2209+1914 & 332.87625 & 19.48690 & 2550, \editas{1200} & 2013 Aug 29, \editas{2016 Sept 21} & \citet{Hagen99} \\
\editas{ATLAS J}2213$-$2652 & 333.41012 & -26.87419 & 420 & 2017 Oct 04 & \citet{Agnello18b} \\
SDSS J2257+2349 & 344.35586 & 23.82510 & 300 & 2016 Sept 22 & \citet{Williams18} \\
PS J2305+3714 & 346.48239 & 37.23899 & 420 & 2017 Oct 04 & \citet{Lemon18} \\
WISE 2329$-$1258 & 352.49114 & -12.98303 & 420 & 2016 Sept 21 & \citet{Schechter17} \\

\hline
\end{tabular}
\end{table*}




We followed up \numobserved\ lens candidates with NIRC2, a near infra-red imager on the W. M. Keck Observatory. These candidates are identified in object catalogues from VST-ATLAS, DES, Hamburg-ESO \editast{\citep[HE;][]{Wisotzki96}}, Pan-STARRS, and SDSS. Searches in the SDSS were based on the population-mixture approach of \citet{Williams17}; searches in Pan-STARRS (PS1) and DES relied on multiplet recognition from the \textit{Gaia}-DR1 catalog \citep{Agnello18c, Lemon18}; and a suite of different methods were applied in the VST-ATLAS searches \citep{Agnello18b}.

We used \editast{the laser guide star adaptive optics (LGS-AO)} whenever \editast{available} \citep{Wizinowich06, vanDam06}\editast{, and observed in seeing-limited mode otherwise}. The images were taken with the $K^{\prime}$ filter on 2016 September, 2017 October, and 2018 January. \editref{The field-of-view (FOV) of the NIRC2 imager is 10$\times$10 squared arcsec in the narrow camera. In this case, the pixel scale is 9.94 mas/pixel. We took an exposure sequence of three 120 s exposures. These three exposures were dithered in a way such that the target lies near the centers of the upper right, upper left, and lower right quadrants of the FOV. We avoided dithering the target into the bad pixel region of the detector in the lower left quadrant of the FOV. In some cases, we also coadded the 60 s exposure -- that was used for target acquisition -- to the final reduced image, if the system is fully contained within the good pixel regions. In ideal circumstances, we aimed for a total exposure time of 1080 s for each lens system, to achieve a sufficiently high signal-to-noise to detect the deflector's light. The total exposure times for each lens system are tabulated in Table \ref{tab:lens_obs}.}

\begin{figure*}
	\includegraphics[width=\textwidth]{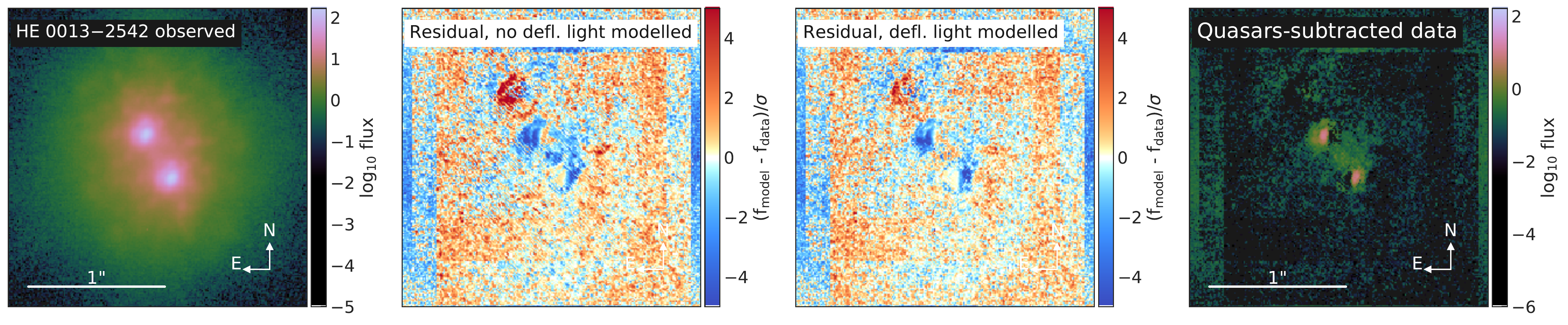}
	\caption{\label{fig:0015_residual} 
	\textbf{First panel:} Observed NIRC2 image of HE 0013$-$2542. \textbf{Second panel:} Noise-normalized residual of the model without the deflector (defl.) light modelled. There are residuals at the quasar image positions from imperfect PSF reconstruction, however there is additional residual in between the quasar image positions indicating the presence of the deflector galaxy. The close proximity of the quasar images combined with irregular shape of the AO PSF makes the PSF reconstruction a difficult task, thus some imperfection in the PSF reconstruction is expected. We detail the modelling and the PSF reconstruction procedures in Section \ref{sec:modelling}. \textbf{Third panel:} Noise-normalized residual of the model with the deflector light modelled. There is no central residual in this case, although the residual at the quasar image positions from imperfect PSF reconstruction persists. As a result, the evidence for the presence of the deflector galaxy is strengthened, as opposed to the hypothesis that the central residual in the second panel is also a result of imperfect PSF reconstruction. \textbf{Fourth panel:} The observed image with the quasar light subtracted using the model from the third panel. Additional light from the deflector galaxy is noticeable in between the two bright points that are remnants from the quasar light subtraction.
	}
\end{figure*}

By visually identifying the presence of a deflector galaxy between the two quasar images, we confirmed 3 lenses out of the 32 observed in 2016, 5 lenses out of the 12 observed in 2017, and 7 lenses out of the 13 observed in 2018. \editref{Thus in total, 15 lenses were confirmed as real lenses through imaging follow-up out of the 57 candidates.} \editas{\editast{For} the remaining 42 candidates, 6 were inconclusive with the rest ruled out as non-lenses.} \editreft{The quasar HE 0013$-$2542
	was previously observed to be a pair at Magellan in
	August 2003, and was re-observed in the following two seasons, but no
	lensing galaxy could be isolated in Sloan $i$ exposures, despite
	excellent seeing.
	} \editreft{In Figure \ref{fig:0015_residual}, we demonstrate that there is additional light in the NIRC2 data in between the quasar image positions that can not be accounted for by the quasar pair. This additional light provides evidence for the presence of the deflector galaxy in between the quasar pair confirming this system as a lensing system.} One of the confirmed lens candidates from the 2018 campaign was later identified as a previously known system\editast{,} LBQS 1009$-$0252. The higher \editast{incidence} of lenses in the later campaigns is due to stricter candidate vetting -- which is also based on false-positive recognition from the 2016 campaign objects -- and due to complementary information from separate, spectroscopic-confirmation campaigns. The full candidate list and outcome of the follow-up imaging is given in Appendix \ref{app:iamgin_campaign}.

\editas{Additionally, we present imaging data from a pilot study carried out in 2013--14 to identify lensed arcs from extended host galaxies in 7 previously known doubles for cosmological applications. These systems were imaged with the NIRC2 $K^{\prime}$ filter in 2013 August and 2014 March (Table \ref{tab:lens_obs}). \editref{These systems were observed for a relatively longer total exposure time (1500--3780 s) to identify the presence of lensed arcs in these systems.}}

Out of the 15 \editas{newly} confirmed systems, two systems had poor image quality in seeing-limited conditions. As a result, the reduction procedure failed to produce science-grade co-added images for them. These two systems are WGA 0146$-$1133 and WGA 0259$-$2338, and they have been \editref{further} confirmed to be lenses through a later spectroscopic follow-up \citep{Agnello18b}. We model the reduced images of the other 13 systems. We add to this sample \editas{the 7 previously known doubles from the 2013--14 campaign}. Thus, the size of our final sample that we model in the next section is 20. We list the coordinates, observing dates, exposure times, and discovery papers for these 20 systems in Table \ref{tab:lens_obs}.

\section{Lens modelling} \label{sec:modelling}
In this section, we describe our lens modelling procedure. We model all the lenses in our sample uniformly. We describe the model components in Section \ref{sec:model_components}, the procedure to estimate the initial point spread function (PSF) in Section \ref{sec:initial_psf}, and the optimization and inference procedure in Section \ref{sec:optimization}.

\subsection{Model components} \label{sec:model_components}
We assume a singular isothermal ellipsoid (SIE) mass profile for the deflector galaxy. The SIE profile is given by
\begin{equation}
	\kappa_{\rm SIE} = \frac{1}{2} \frac{\theta_{\rm E}}{\sqrt{q_{\rm m}\theta_1^2 + \theta_2^2/q_{\rm m}}},
\end{equation}
where $\theta_{\rm E}$ is the Einstein radius and $q_{\rm m}$ is the axis ratio. The coordinates $(\theta_1, \theta_2)$ are rotated by position angle PA$_{\rm m}$ relative to the on-sky coordinate systems (RA, dec) \editast{to align with the major axis of the projected mass distribution}. 

We model the light distribution in the deflector galaxy with elliptical de Vaucouleurs' profile \citep{deVaucouleurs48}. We parameterize the axis ratio of the deflector light distribution with $q_{\rm L}$ and the position angle with PA$_{\rm L}$. For two lenses, PS J0417+3325 and SDSS J2257+2349, the deflector galaxy's light distribution fits poorly to the de Vaucouleurs' profile. As these two galaxies appear disky, we adopt an additional exponential profile to model the disk component. For simplicity, we take the exponential profile and the de Vaucouleurs' profile to have the same axis ratio $q_{\rm L}$ and position angle PA$_{\rm L}$. \editreft{Only for HE 0013$-$2542, we adopt a circular de Vaucouleurs' profile, because an elliptical de Vaucouleurs' profile makes the lens model unstable due to poor constraints as the deflector galaxy's light is vastly overshadowed by the quasar images in close proximity.}

\begin{figure}
	\centering
	\includegraphics[width=\columnwidth]{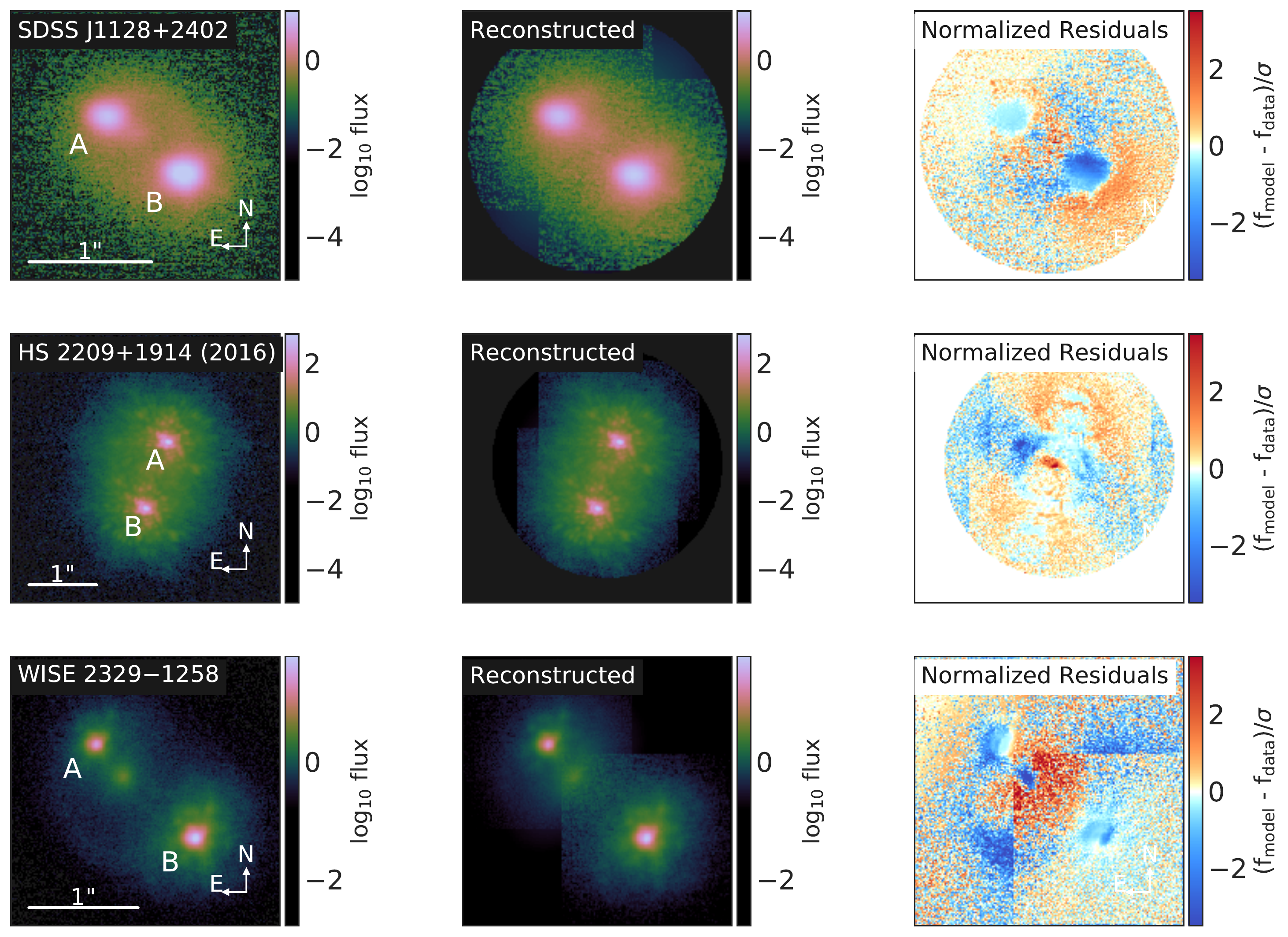}
	\caption{\label{fig:wo_source}
		\editref{Lens models of the systems HS 2209$+$1914, SDSS J1128$+$2402, and WISE 2329$-$1258 without including the quasar host galaxy light in the model. The image residuals show arc-like negative (blue) residuals, which is indicative of the faint lensed arcs from the host galaxy. Therefore, we include the quasar host galaxy light profile in the model. The final models for these lens systems are illustrated in Figure \ref{fig:montage_2}.}
	}
\end{figure}

We treat the lensed quasar images as point sources and \editast{we \editref{model them} on the image plane}. \editas{The} HE 0047$-$1756 and SDSS J0246$-$0825 \editast{systems} \editas{have} prominent arcs. For some other systems, the fit was poor when the host galaxy light was not explicitly modelled at first, and residuals from the lensed arcs were noticeable by eye in the difference between the data and the model-based-reconstruction \editref{(Figure \ref{fig:wo_source})}. For these systems, we adopt an elliptical S\'ersic profile \editas{on the source plane} to capture the extended light distribution of the quasar host galaxy \citep{Sersic68}. These systems are HS 2209+1914, SDSS J1128+2402, and WISE 2329$-$1258. \editref{In these systems, the lens models are constrained from the image positions, the deflector centroid, and the lensed arcs. For the remaining systems, the lens model is only constrained from the image positions and the deflector centroid.}

\subsection{Estimation of initial PSF} \label{sec:initial_psf}

\begin{figure}
	\centering
	\includegraphics[width=0.8\columnwidth]{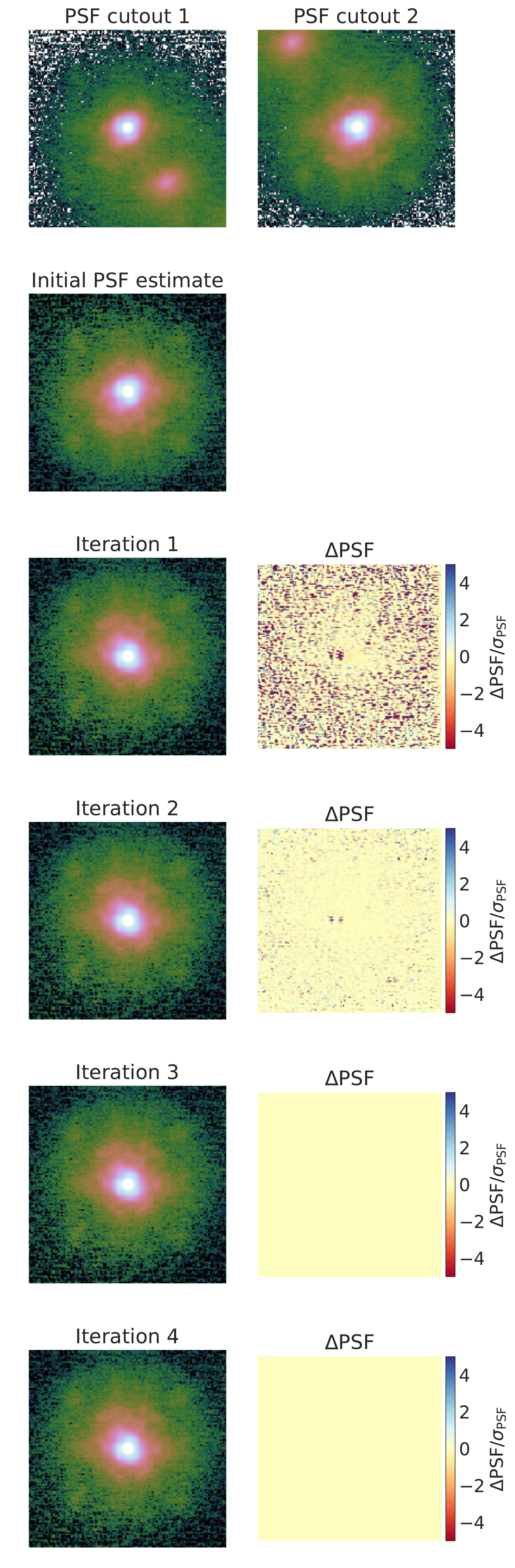}
	\caption{\label{fig:psf_iteration}
		{Estimation of the initial PSF and PSF reconstruction through an iterative algorithm. This examples is shown for the system PS J0140$+$4107. The top row shows two cutouts centered at the quasar images. The contamination in the cutouts are removed through a weighted sum of the two cutouts to obtain the initial PSF estimate in the second row. The third to last rows show the reconstructed PSF through an iterative procedure (first column) described in Section \ref{sec:optimization} and the difference between the current PSF iteration with the previous one (second column). The reconstructed PSF typically converges after 2--3 iterations.}
	}
\end{figure}

An accurate PSF is necessary to model the quasar images with point sources. To reconstruct the PSF for each lens, we first estimate an initial PSF that we then iteratively optimize (as we will describe in Section \ref{sec:optimization}). Given the limited field of view of the NIRC2 imager, we do not have nearby stars within the observed image to use as the initial PSF estimate. Although the lensed quasars themselves are point sources, often times their light distributions are blended with the deflector galaxy's light \editas{or with the lensed arcs from the quasar host galaxy}. We \editas{adopt} the following strategy to minimize the contamination of the deflector's light in the initial PSF estimated from the two quasar images. We first take the cutouts of two quasar images. These cutouts often include the extended light from the deflector. As the two quasar images do not have the same magnitude, we scale up the the fainter quasar image's cutout so that the two cutouts have the same peak value. At the pixel with the peak value within each cutout, the contamination fraction from the deflector is minimum. However, the contamination fraction from the deflector light would increase towards the deflector's position. For each pair of corresponding pixels between the two cutouts, we take a weighted average of the two pixels with weight 1 for the lower value and weight $\exp\left(-\Delta I^2/(\sigma_1^2 + \sigma_2^2)\right)$ for the higher value, where $\Delta I$ is the difference between the the pixel values between corresponding pixels, and $\sigma_1$, $\sigma_2$ are noise levels for the two pixels. \editref{This weighting scheme assumes that the pixel with the lower value within the pair is the ``reliable'' measurement and then weights the other pixel value by its probability under the probability distribution of the ``reliable'' measurement.} We use this weighted average of the two quasar image cutouts as the initial PSF estimate. The noise map for the initial estimate is also obtained considering the same weights while combining noise from the two cutouts. As we will eventually reconstruct the PSF by iterative optimization, this initial estimate only needs to be close enough to the truth so that the iterative optimization can successfully converge. \editref{The first two rows of Figure \ref{fig:psf_iteration} demonstrate an example of this algorithm to estimate the initial PSF from two cutouts with contaminants.}

For the HE 0047$-$1756 and SDSS 0246$-$0825 \editast{systems}, the contamination from the  prominent lensed arcs \editas{can} not be sufficiently minimized using the above strategy to allow for a successful reconstruction. For that reason, we use the reconstructed PSF from SDSS J1001+5027 as the initial PSF estimate for these two systems, which leads to a better model fit; although residuals are still noticeable \editref{in the difference between the data and the model-based-reconstruction}. \editas{We choose the reconstructed PSF from SDSS J1001+5027 due to its relatively smooth profile and nearly circular shape, which are preferable features in an initial PSF estimate when a more reliable one is lacking.}

\subsection{Optimization and inference} \label{sec:optimization}

We model the lenses with the lens modelling software \textsc{lenstronomy} \editas{\citep{Birrer15, Birrer18}}. \editas{\textsc{lenstronomy} is an open-source software available online at GitHub.\footnote{\faGithub\ \url{https://github.com/sibirrer/lenstronomy}}} We first iteratively reconstruct the PSF by alternatively optimizing the lens model and the initial PSF estimate \editas{\citep{Birrer16, Chen16, Shajib19, Birrer19}}. \editref{For each iteration, the lens model is first optimized using the currently estimated PSF. Then, the PSF is optimized by subtracting the modelled deflector light (and the lensed lensed quasar host if in the model), and then minimizing the image residuals around the quasar image positions.} After \editref{2--3 such iterations}, the image likelihood does not increase with further iterations of the PSF reconstruction. Therefore, we \editref{take four such iterations to be sufficient for reliable convergence of the PSF reconstruction}. \editref{The bottom four rows of Figure \ref{fig:psf_iteration} demonstrate an example of the reconstructed PSF at each iteration.}

We optimize the lens model during the PSF reconstruction using particle swarm optimization \citep[PSO;][]{Kennedy95}. After the PSF reconstruction, we execute a Markov chain Monte Carlo (MCMC) using \textsc{emcee} to obtain the posterior probability distributions of the model parameters \citep{Goodman10, Foreman-Mackey13}. We confirm \editas{the convergence of the} MCMC chain by checking that the median and the standard deviation of the \textsc{emcee} walkers at each step have stabilized for $\mathcal{O}(10)$ times the autocorrelation length \editas{\citep{Foreman-Mackey13}}. 

The HS 2209+1914 \editast{system was} imaged in two different campaigns. To model this system, we simultaneously use images from both 2013 and 2016 with \editas{separately reconstructed} PSF for each. 

\begin{figure*}
	\includegraphics[width=\textwidth]{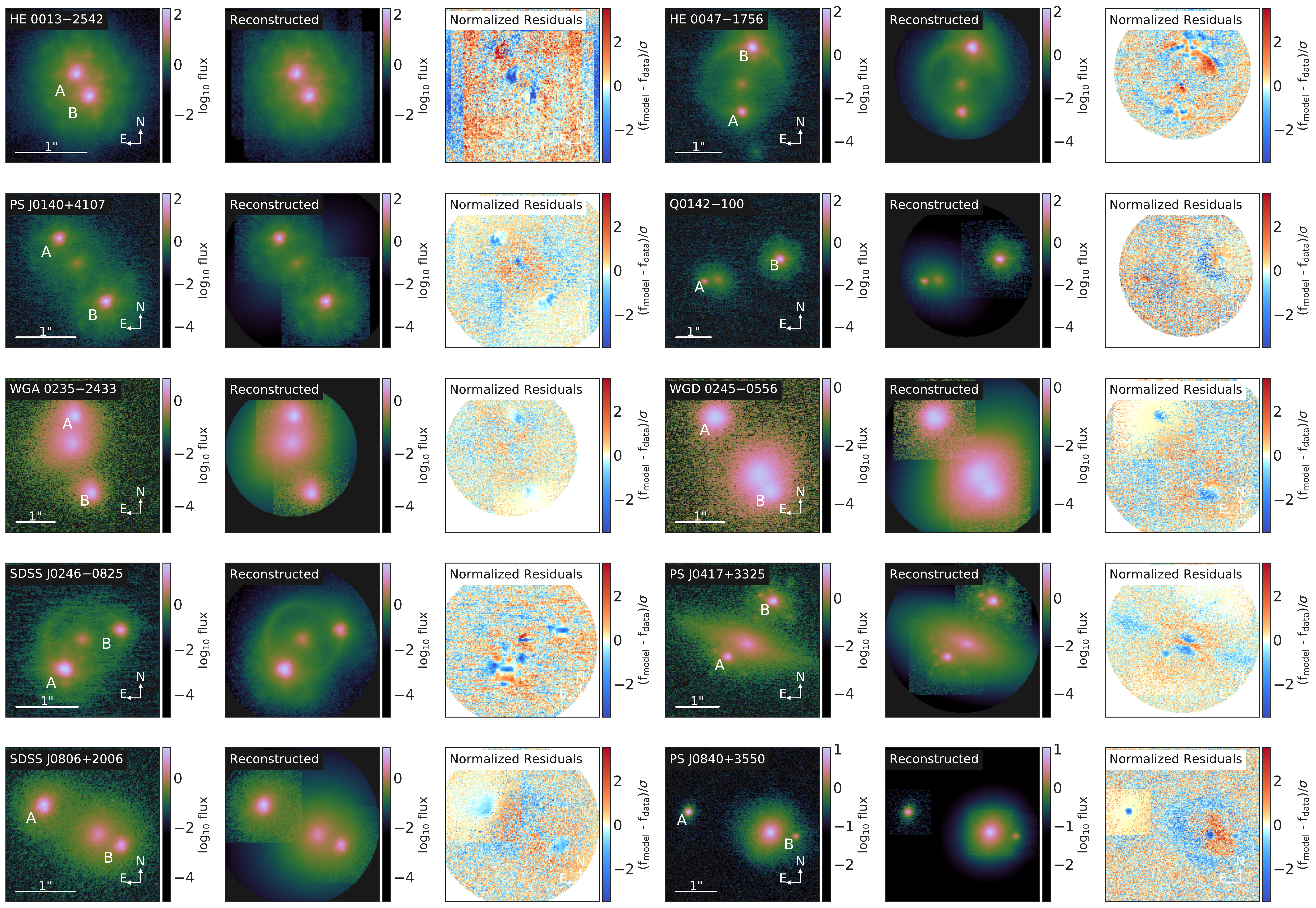}
	\caption{\label{fig:montage_1}
		Observed image and model for the first 10 doubles. The first and the fourth columns show the observed Keck/NIRC2 images of the lens systems. The second and the fifth columns show the models. The third and the sixth columns show the noise-normalized residuals. The whitened out areas in the residual maps are due to our adopted masks.
	}
\end{figure*}

\begin{figure*}
	\includegraphics[width=\textwidth]{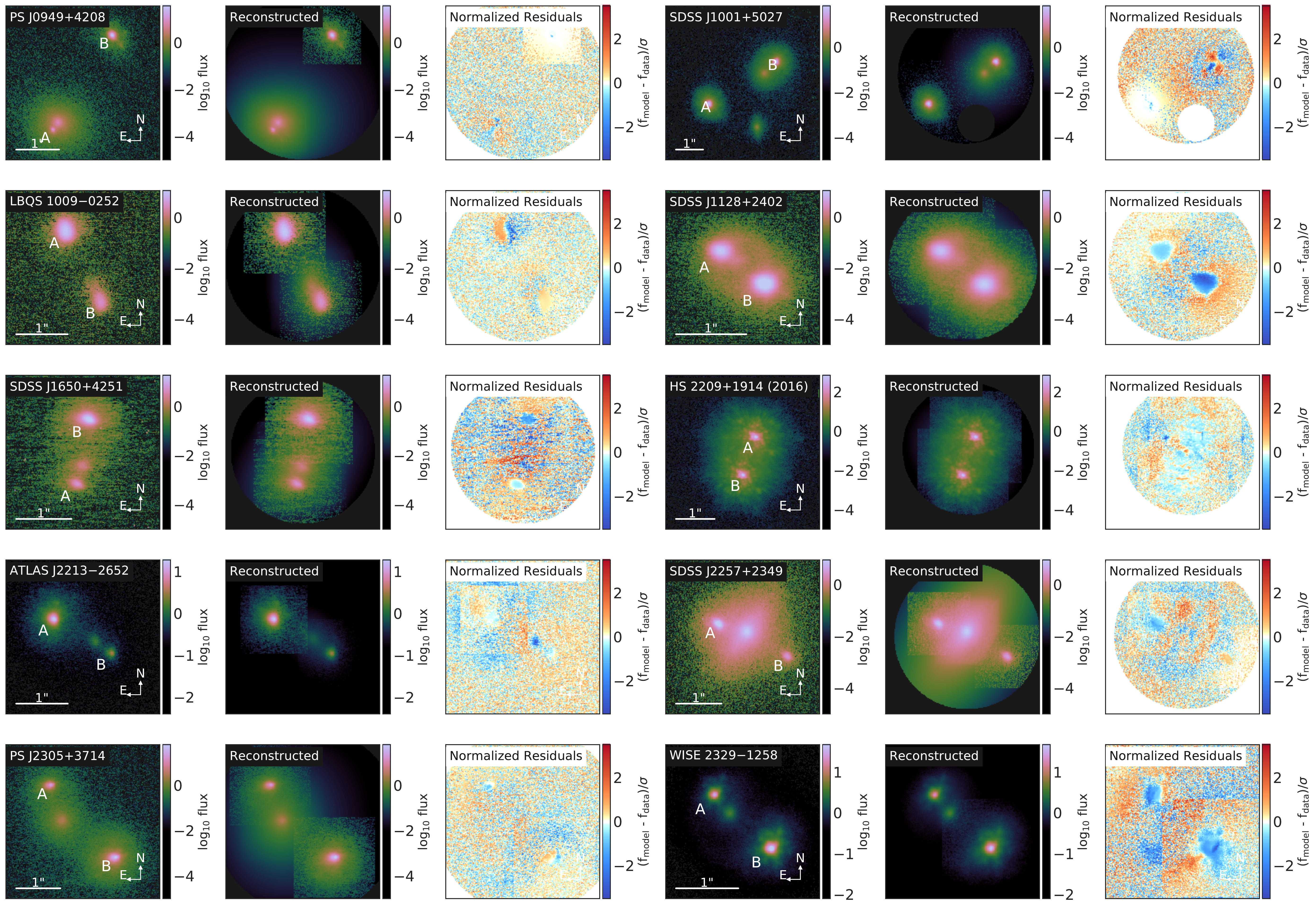}
	\caption{\label{fig:montage_2}
		Continuation of Figure \ref{fig:montage_1} for the last 10 doubles. We show only the 2016 image of HS2209+1914. However, both images from 2013 and 2016 were simultaneously used to model this system.
	}
\end{figure*}

We compare the model-reconstructions with the observed images for all \nummodeled\ systems in Figures \ref{fig:montage_1} and \ref{fig:montage_2}. \editas{The irregularity of the AO PSF in our images makes it difficult to accurately reconstruct the PSF. As a result, prominent residuals in Figures \ref{fig:montage_1} and \ref{fig:montage_2} are noticeable, specially around the quasar image positions. A more accurate PSF reconstruction similar to \citet{Chen16, Chen19} would require careful treatment on a lens-by-lens basis. In this paper, we focus on uniform modeling of a large sample, thus a lens-by-lens treatment of the PSF reconstruction is beyond the scope and requirement of this paper.}


\section{Astrometry, photometry, and model parameters} \label{sec:results}

From the lens models, we provide \editref{relative astrometry and relative photometry} of the deflector galaxy and the quasar images in Table \ref{tab:astrometry_photometry}. \editref{The initial PSF estimate in our modelling is not centered within the central pixel with the peak value. This can lead to a potential systematic error, if the model is over-optimized to the initial PSF estimate in the first iteration of the PSF reconstruction process. Therefore, we add a systematic uncertainty of 0.005 arcsec -- which is approximately half the pixel size -- to the statistical uncertainty of the astrometric positions in quadrature.} We obtain the total flux of the deflector galaxy by \editref{analytically} integrating the \editref{modelled surface brightness} profile up to infinity. \editrefs{We use the NIRC2 zero-point magnitude $m_0=24.74$ to convert the total flux into apparent magnitude.\footnote{\url{https://www2.keck.hawaii.edu/inst/nirc2/filters.html}} We add $\sigma_{m_{0}}=0.025$ uncertainty in quadrature to the statistical uncertainty to account for the typical error in the zero-point magnitude correction \citep[][see figure 19 therein]{Gautam19}. \editast{Note, we generally did not observe in photometric condition. Thus, the reported magnitudes can potentially be affected by atmospheric extinction and instrumental transmission. We are unable to correct for these extinction effects as NIRC2's narrow field-of-view did not allow us to simultaneously observe a standard star for photometric calibration.}}


We tabulate the Einstein radius, the effective radius, and the ellipticities and position angles for both mass and light distributions in Table \ref{tab:lens_models}. For the lenses with \editref{de Vaucouleurs' + exponential} profile fits, we numerically compute the half-light radius as the effective radius. We also provide the Fermat potential difference between the quasar images in Table \ref{tab:lens_models}. The Fermat potential difference is given by
\begin{equation}
	\Delta \phi_{\rm AB} = \frac{\left(\btheta_{\rm A} - \bbeta\right)^2}{2} - \frac{\left(\btheta_{\rm B} - \bbeta\right)^2}{2} - \psi(A) + \psi(B),
\end{equation}
where $\btheta$ is the image position, $\bbeta$ is the source position, and $\psi$ is the deflection potential. If the redshifts of the source and the deflector are known, then the time-delay $\Delta t_{\rm AB}$ between the images can be computed for a given cosmology as
\begin{equation}
	\Delta t_{\rm AB} = \frac{(1+z_{\rm d})}{c} \frac{D_{\rm d} D_{\rm s}}{D_{\rm ds}} \Delta \phi_{\rm AB}.	
\end{equation}
Here, $c$ is the speed of the light and $z_{\rm d}$ is the deflector redshift. The angular diameter distances are $D_{\rm d}$: between the observer and the deflector, $D_{\rm s}$: between the observer and the source, and $D_{\rm ds}$: between the deflector and the source. \editas{For quick reference, we also provide the corresponding time-delays in unit of days assuming fiducial redshifts $z_{\rm d}=0.5$ for the deflector, $z_{\rm s}=2$ for the source, and a fiducial flat $\Lambda$CDM cosmology with $h=0.7$, $\Omega_{\rm m}=0.3$ (Table \ref{tab:lens_models}).}

\renewcommand{\arraystretch}{1.2}
\begin{table*}
\caption{\label{tab:astrometry_photometry}
	Astrometry and photometry of the lens galaxy and quasar images. $\Delta$RA and $\Delta$dec of the images are computed taking (RA, dec)${}_{\rm G}\equiv(0,0)$. \editref{A systematic uncertainty of 0.005 arcsec is added in quadrature to the astrometric positions to account for potential systematic due to not centering the initial PSF estimate within the central peak pixel.} \editref{The magnitude of the deflector galaxy is computed from integrating the modelled surface brightness profile up to infinity.}
}
\begin{tabular}{lccccccc}
\hline
Name & $\Delta$RA$_{\rm A}$ & $\Delta$dec$_{\rm A}$ & $\Delta$RA$_{\rm B}$ &$\Delta$dec$_{\rm B}$ & $m_{\rm A}$ & $m_{\rm B} - m_{\rm A}$ & $m_{\rm G} - m_{\rm A}$ \\
 & (arcsec) & (arcsec) & (arcsec) & (arcsec) & (mag) & (mag) & (mag) \\
\hline

HE 0013$-$2542 & $0.116_{-0.003 }^{ +0.002 }$ & $0.162_{-0.003 }^{ +0.003 }$ & $-0.065_{-0.003 }^{ +0.002 }$ & $-0.155_{-0.003 }^{ +0.003 }$ & $13.18 \pm 0.03$ & $-0.0715 \pm 0.0004$ & $4.61 \pm 0.03$ \\
HE 0047$-$1756 & $-0.0031_{-0.0005 }^{ +0.0004 }$ & $-0.6014_{-0.0084 }^{ +0.0004 }$ & $-0.2336_{-0.0001 }^{ +0.0001 }$ & $0.8108_{-0.0101 }^{ +0.0001 }$ & $15.10 \pm 0.04$ & $-1.69 \pm 0.02$ & $-0.05 \pm 0.08$ \\
PS J0140+4107 & $0.2908_{-0.0002 }^{ +0.0003 }$ & $0.4260_{-0.0100 }^{ +0.0001 }$ & $-0.4897_{-0.0001 }^{ +0.0002 }$ & $-0.6340_{-0.0098 }^{ +0.0001 }$ & $14.60 \pm 0.03$ & $-0.597 \pm 0.001$ & $0.61 \pm 0.04$ \\
Q0142$-$100 & $0.3813_{-0.0003 }^{ +0.0003 }$ & $-0.0380_{-0.0003 }^{ +0.0003 }$ & $-1.7535_{-0.0003 }^{ +0.0003 }$ & $0.5815_{-0.0003 }^{ +0.0002 }$ & $15.59 \pm 0.03$ & $-2.098 \pm 0.003$ & $-0.90 \pm 0.01$ \\
WGA 0235$-$2433 & $-0.041_{-0.001 }^{ +0.001 }$ & $0.698_{-0.001 }^{ +0.001 }$ & $-0.485_{-0.001 }^{ +0.001 }$ & $-1.287_{-0.001 }^{ +0.001 }$ & $14.87 \pm 0.03$ & $0.312 \pm 0.002$ & $-1.01 \pm 0.01$ \\
WGD 0245$-$0556 & $0.9145_{-0.0003 }^{ +0.0004 }$ & $1.1733_{-0.0002 }^{ +0.0001 }$ & $-0.2359_{-0.0002 }^{ +0.0002 }$ & $-0.3009_{-0.0003 }^{ +0.0002 }$ & $15.63 \pm 0.03$ & $0.377 \pm 0.003$ & $-0.99 \pm 0.01$ \\
SDSS J0246$-$0825 & $0.2842_{-0.0002 }^{ +0.0003 }$ & $-0.4814_{-0.0002 }^{ +0.0004 }$ & $-0.6117_{-0.0004 }^{ +0.0017 }$ & $0.1486_{-0.0008 }^{ +0.0004 }$ & $14.14 \pm 0.03$ & $1.79 \pm 0.02$ & $1.77 \pm 0.03$ \\
PS J0417+3325 & $0.4536_{-0.0004 }^{ +0.0004 }$ & $-0.2972_{-0.0002 }^{ +0.0003 }$ & $-0.6029_{-0.0003 }^{ +0.0003 }$ & $0.9917_{-0.0002 }^{ +0.0003 }$ & $15.58 \pm 0.03$ & $-0.577 \pm 0.002$ & $-1.732 \pm 0.005$ \\
SDSS J0806+2006 & $0.9259_{-0.0001 }^{ +0.0002 }$ & $0.4982_{-0.0025 }^{ +0.0001 }$ & $-0.3888_{-0.0002 }^{ +0.0003 }$ & $-0.1779_{-0.0004 }^{ +0.0003 }$ & $15.44 \pm 0.03$ & $0.969 \pm 0.002$ & $-0.51 \pm 0.01$ \\
PS J0840+3550 & $2.02818_{-0.00002 }^{ +0.00002 }$ & $0.49711_{-0.00001 }^{ +0.00004 }$ & $-0.6305_{-0.0004 }^{ +0.0004 }$ & $-0.1017_{-0.0004 }^{ +0.0005 }$ & $16.40 \pm 0.03$ & $1.18 \pm 0.01$ & $-2.655 \pm 0.002$ \\
PS J0949+4208 & $0.1126_{-0.0004 }^{ +0.0005 }$ & $-0.1667_{-0.0005 }^{ +0.0004 }$ & $-1.2572_{-0.0003 }^{ +0.0003 }$ & $2.0253_{-0.0003 }^{ +0.0003 }$ & $17.06 \pm 0.03$ & $-1.63 \pm 0.01$ & $-2.95 \pm 0.01$ \\
SDSS J1001+5027 & $2.0403_{-0.0002 }^{ +0.0002 }$ & $-1.1206_{-0.0002 }^{ +0.0003 }$ & $-0.4223_{-0.0002 }^{ +0.0002 }$ & $0.4331_{-0.0002 }^{ +0.0002 }$ & $14.17 \pm 0.03$ & $0.112 \pm 0.001$ & $0.626 \pm 0.003$ \\
LBQS 1009$-$0252 & $0.54_{-0.01 }^{ +0.01 }$ & $1.11_{-0.01 }^{ +0.01 }$ & $-0.16_{-0.01 }^{ +0.01 }$ & $-0.25_{-0.01 }^{ +0.01 }$ & $14.81 \pm 0.03$ & $1.37 \pm 0.01$ & $1.77 \pm 0.15$ \\
SDSS J1128+2402 & $0.229_{-0.002 }^{ +0.003 }$ & $0.147_{-0.004 }^{ +0.003 }$ & $-0.385_{-0.002 }^{ +0.003 }$ & $-0.327_{-0.005 }^{ +0.003 }$ & $15.56 \pm 0.03$ & $-0.16 \pm 0.02$ & $1.38 \pm 0.12$ \\
SDSS J1650+4251 & $0.08947_{-0.00005 }^{ +0.01010 }$ & $-0.30819_{-0.00031 }^{ +0.00003 }$ & $-0.1226_{-0.0001 }^{ +0.0100 }$ & $0.8519_{-0.0002 }^{ +0.0001 }$ & $16.10 \pm 0.06$ & $-1.45 \pm 0.03$ & $-0.50 \pm 0.33$ \\
HS 2209+1914 & $-0.227_{-0.001 }^{ +0.001 }$ & $0.387_{-0.001 }^{ +0.001 }$ & $0.099_{-0.001 }^{ +0.001 }$ & $-0.599_{-0.001 }^{ +0.001 }$ & $12.43 \pm 0.03$ & $0.139 \pm 0.001$ & $3.30 \pm 0.04$ \\
A2213$-$2652 & $0.764_{-0.002 }^{ +0.008 }$ & $0.3842_{-0.0004 }^{ +0.0097 }$ & $-0.356_{-0.002 }^{ +0.008 }$ & $-0.2783_{-0.0003 }^{ +0.0099 }$ & $14.68 \pm 0.03$ & $2.12 \pm 0.02$ & $-0.63 \pm 0.10$ \\
SDSS J2257+2349 & $0.6194_{-0.0002 }^{ +0.0002 }$ & $0.1753_{-0.0003 }^{ +0.0003 }$ & $-0.8792_{-0.0002 }^{ +0.0002 }$ & $-0.5433_{-0.0003 }^{ +0.0003 }$ & $15.23 \pm 0.03$ & $0.945 \pm 0.003$ & $-1.96 \pm 0.01$ \\
PS J2305+3714 & $0.267_{-0.001 }^{ +0.001 }$ & $0.793_{-0.001 }^{ +0.001 }$ & $-1.182_{-0.001 }^{ +0.001 }$ & $-0.835_{-0.001 }^{ +0.001 }$ & $15.26 \pm 0.03$ & $-1.204 \pm 0.002$ & $-0.98 \pm 0.01$ \\
WISE 2329$-$1258 & $0.2411_{-0.0004 }^{ +0.0003 }$ & $0.298_{-0.001 }^{ +0.001 }$ & $-0.6557_{-0.0004 }^{ +0.0003 }$ & $-0.555_{-0.002 }^{ +0.001 }$ & $15.34 \pm 0.03$ & $-1.00 \pm 0.01$ & $0.56 \pm 0.01$ \\

\hline
\end{tabular}
\end{table*}

\renewcommand{\arraystretch}{1.2}
\begin{table*}
\caption{\label{tab:lens_models}
Model parameters and estimated Fermat potential difference $\Delta \phi_{\rm AB}$ for the lens systems. \editref{A conservative systematic uncertainty of 0.01 arcsec is added in quadrature to the statistical uncertainty of the Einstein radius $\theta_{\rm E}$ to account for potential systematic caused by an uncentered initial PSF estimate.} \editas{The position angles PA$_{\rm m}$ and PA$_{\rm L}$ are \editref{North of East}. 
\editrefs{We restrict the reference axis for PA within a particular quadrant and allow the axis ratio $q$ to be $>1$, which corresponds to case when the PA refers to the orientation of the minor axis. This convention helps to avoid bi-modality in the PA distributions for some systems.}
\editreft{The HE 0013$-$2542 system is modelled with a circular deflector light profile, thus no axis ratio for the deflector light is provided. Furthermore, we only provide a 95 per cent upper limit on the effective radius.} The reported time delays $\Delta t_{\rm AB}$ are computed using fiducial redshifts $z_{\rm d}=0.5$ for the deflector, and $z_{\rm s}=2$ for the source, and for a fiducial flat $\Lambda$CDM cosmology with $h=0.7$, $\Omega_{\rm m}=0.3$.} \editast{Positive $\Delta t_{\rm AB}$ implies that image B leads image A, and negative value implies the opposite.}
}
\begin{tabular}{lcccccccc}
\hline
       Name & $\theta_{\rm E}$ & $q_{\rm m}$ & PA$_{\rm m}$ & $\theta_{\rm eff}$ & $q_{\rm L}$ & PA$_{\rm L}$ & $\Delta \phi_{\rm AB}$ & $\Delta t_{\rm AB}$ \\
       & (arcsec) & & (deg) & (arcsec) & & (deg) & ($\times10^{-13}$) & (d)\\
\hline

HE 0013$-$2542 & $0.18_{-0.01 }^{ +0.04 }$ & $1.13_{-0.63 }^{ +1.04 }$ & $60_{-2 }^{ +3 }$ & $<0.103$ & -- & -- & $-1.34_{-0.30 }^{ +0.24 }$ & $-0.47_{-0.11 }^{ +0.08 }$ \\
HE 0047$-$1756 & $0.745_{-0.001 }^{ +0.001 }$ & $0.840_{-0.008 }^{ +0.003 }$ & $86.3_{-0.3 }^{ +1.4 }$ & $0.50_{-0.06 }^{ +0.01 }$ & $0.77_{-0.01 }^{ +0.01 }$ & $-38_{-9 }^{ +2 }$ & $41.16_{-3.09 }^{ +0.06 }$ & $14.58_{-1.09 }^{ +0.02 }$ \\
PS J0140$+$4107 & $0.65_{-0.04 }^{ +0.04 }$ & $1.06_{-0.29 }^{ +1.23 }$ & $53_{-1 }^{ +6 }$ & $0.67_{-0.07 }^{ +0.01 }$ & $0.56_{-0.01 }^{ +0.01 }$ & $-32_{-1 }^{ +1 }$ & $44.16_{-0.03 }^{ +0.63 }$ & $15.64_{-0.01 }^{ +0.22 }$ \\
Q0142$-$100 & $1.26_{-0.08 }^{ +0.13 }$ & $0.55_{-0.15 }^{ +0.14 }$ & $-3_{-4 }^{ +9 }$ & $0.453_{-0.003 }^{ +0.004 }$ & $0.748_{-0.007 }^{ +0.004 }$ & $34_{-1 }^{ +1 }$ & $383.80_{-0.17 }^{ +0.14 }$ & $135.92_{-0.06 }^{ +0.05 }$ \\
WGA 0235$-$2433 & $0.97_{-0.02 }^{ +0.04 }$ & $0.54_{-0.15 }^{ +0.18 }$ & $-22_{-11 }^{ +4 }$ & $0.52_{-0.01 }^{ +0.01 }$ & $0.72_{-0.01 }^{ +0.01 }$ & $-40_{-1 }^{ +1 }$ & $164.85_{-0.37 }^{ +0.32 }$ & $58.38_{-0.13 }^{ +0.11 }$ \\
WGD 0245$-$0556 & $0.89_{-0.03 }^{ +0.09 }$ & $1.41_{-0.62 }^{ +0.87 }$ & $52.1_{-0.4 }^{ +0.3 }$ & $0.42_{-0.01 }^{ +0.01 }$ & $0.63_{-0.01 }^{ +0.01 }$ & $-52_{-1 }^{ +1 }$ & $-242.86_{-0.11 }^{ +0.10 }$ & $-86.00_{-0.04 }^{ +0.03 }$ \\
SDSS J0246$-$0825 & $0.5877_{-0.0009 }^{ +0.0005 }$ & $0.901_{-0.002 }^{ +0.002 }$ & $83_{-1 }^{ +1 }$ & $0.26_{-0.01 }^{ +0.01 }$ & $0.89_{-0.02 }^{ +0.42 }$ & $-28_{-6 }^{ +13 }$ & $9.83_{-0.10 }^{ +0.09 }$ & $3.48_{-0.04 }^{ +0.03 }$ \\
PS J0417$+$3325 & $0.80_{-0.01 }^{ +0.02 }$ & $0.52_{-0.13 }^{ +0.16 }$ & $29_{-8 }^{ +3 }$ & $0.335_{-0.002 }^{ +0.002 }$ & $0.349_{-0.003 }^{ +0.002 }$ & $18.0_{-0.1 }^{ +0.1 }$ & $123.73_{-0.10 }^{ +0.08 }$ & $43.82_{-0.04 }^{ +0.03 }$ \\
SDSS J0806$+$2006 & $0.76_{-0.07 }^{ +0.10 }$ & $0.85_{-0.35 }^{ +1.09 }$ & $26_{-3 }^{ +4 }$ & $0.44_{-0.01 }^{ +0.01 }$ & $0.95_{-0.01 }^{ +0.01 }$ & $-76_{-3 }^{ +4 }$ & $-108.41_{-0.04 }^{ +0.28 }$ & $-38.39_{-0.01 }^{ +0.10 }$ \\
PS J0840$+$3550 & $1.39_{-0.12 }^{ +0.17 }$ & $1.08_{-0.54 }^{ +0.94 }$ & $14_{-9 }^{ +5 }$ & $0.420_{-0.002 }^{ +0.001 }$ & $0.929_{-0.003 }^{ +0.003 }$ & $-31_{-1 }^{ +2 }$ & $-464.51_{-0.07 }^{ +0.06 }$ & $-164.50_{-0.03 }^{ +0.02 }$ \\
PS J0949$+$4208 & $1.25_{-0.05 }^{ +0.19 }$ & $0.79_{-0.38 }^{ +0.86 }$ & $31_{-2 }^{ +6 }$ & $0.544_{-0.003 }^{ +0.003 }$ & $0.831_{-0.004 }^{ +0.004 }$ & $-16_{-1 }^{ +1 }$ & $662.96_{-0.17 }^{ +0.20 }$ & $234.78_{-0.06 }^{ +0.07 }$ \\
SDSS J1001$+$5027 & $1.39_{-0.03 }^{ +0.04 }$ & $0.59_{-0.17 }^{ +0.13 }$ & $70_{-5 }^{ +11 }$ & $0.686_{-0.003 }^{ +0.003 }$ & $0.845_{-0.003 }^{ +0.004 }$ & $-85_{-1 }^{ +1 }$ & $-593.77_{-0.17 }^{ +0.19 }$ & $-210.27_{-0.06 }^{ +0.07 }$ \\
LBQS 1009$-$0252 & $0.74_{-0.03 }^{ +0.11 }$ & $1.34_{-0.72 }^{ +0.96 }$ & $66_{-12 }^{ +9 }$ & $0.40_{-0.07 }^{ +0.09 }$ & $2.00_{-1.20 }^{ +0.12 }$ & $-9_{-50 }^{ +4 }$ & $-166.67_{-3.45 }^{ +2.87 }$ & $-59.02_{-1.22 }^{ +1.02 }$ \\
SDSS J1128$+$2402 & $0.394_{-0.001 }^{ +0.001 }$ & $0.91_{-0.01 }^{ +0.01 }$ & $16_{-2 }^{ +2 }$ & $0.35_{-0.05 }^{ +0.06 }$ & $0.66_{-0.03 }^{ +0.02 }$ & $-52_{-8 }^{ +11 }$ & $21.28_{-0.55 }^{ +0.68 }$ & $7.54_{-0.20 }^{ +0.24 }$ \\
SDSS J1650$+$4251 & $0.580_{-0.004 }^{ +0.004 }$ & $0.84_{-0.03 }^{ +0.05 }$ & $-9_{-6 }^{ +5 }$ & $0.35_{-0.03 }^{ +0.24 }$ & $0.53_{-0.02 }^{ +0.05 }$ & $-42_{-6 }^{ +1 }$ & $74.95_{-0.67 }^{ +0.02 }$ & $26.54_{-0.24 }^{ +0.01 }$ \\
HS 2209$+$1914 & $0.505_{-0.001 }^{ +0.001 }$ & $0.73_{-0.01 }^{ +0.01 }$ & $10.9_{-0.3 }^{ +0.3 }$ & $0.17_{-0.01 }^{ +0.01 }$ & $0.64_{-0.02 }^{ +0.03 }$ & $30_{-3 }^{ +3 }$ & $19.62_{-0.20 }^{ +0.21 }$ & $6.95 \pm 0.07$ \\
A2213$-$2652 & $0.63_{-0.03 }^{ +0.04 }$ & $1.32_{-0.46 }^{ +0.97 }$ & $27_{-11 }^{ +11 }$ & $3.00_{-0.53 }^{ +0.18 }$ & $0.68_{-0.02 }^{ +0.03 }$ & $49_{-2 }^{ +3 }$ & $-61.98_{-3.56 }^{ +0.61 }$ & $-21.95_{-1.26 }^{ +0.22 }$ \\
SDSS J2257$+$2349 & $0.85_{-0.07 }^{ +0.13 }$ & $1.11_{-0.60 }^{ +0.89 }$ & $28_{-10 }^{ +6 }$ & $0.41_{-0.01 }^{ +0.01 }$ & $0.360_{-0.003 }^{ +0.003 }$ & $-56.6_{-0.2 }^{ +0.1 }$ & $76.84_{-0.09 }^{ +0.08 }$ & $27.21 \pm 0.03$ \\
PS J2305$+$3714 & $1.26_{-0.22 }^{ +0.13 }$ & $0.62_{-0.15 }^{ +1.96 }$ & $64_{-22 }^{ +13 }$ & $0.77_{-0.01 }^{ +0.01 }$ & $0.77_{-0.01 }^{ +0.01 }$ & $79_{-1 }^{ +1 }$ & $163.77_{-0.49 }^{ +0.42 }$ & $58.00_{-0.18 }^{ +0.15 }$ \\
WISE 2329$-$1258 & $0.618_{-0.001 }^{ +0.001 }$ & $0.889_{-0.003 }^{ +0.002 }$ & $-80_{-2 }^{ +2 }$ & $0.123_{-0.003 }^{ +0.005 }$ & $0.67_{-0.02 }^{ +0.04 }$ & $-61_{-4 }^{ +1 }$ & $69.54_{-0.13 }^{ +0.59 }$ & $24.63_{-0.05 }^{ +0.21 }$ \\

\hline
\end{tabular}	
\end{table*}

\section{Discussion} \label{sec:discussion}
In this paper, we present lens models of \nummodeled\ doubly imaged quasar systems. 13 of these systems were imaged with NIRC2 in 2016--2018 and confirmed as lenses from a pool of \numobserved\ observed candidates. \editas{The other 7 systems were imaged as part of a pilot program to identify doubles with extended arcs for cosmological applications.} From their lens models, we provide astrometry and photometry of the deflector galaxies and the quasar images. We also present the estimated lens model parameters -- e.g., Einstein radii and effective radii -- and the Fermat potential differences between the images. \editas{This} information will facilitate planning of future follow-up observations to gather ancillary data for various astrophysical applications. \editref{We also report on a new lens system HE 0013$-$2542 for the first time in the literature.}

\begin{figure*}
	\includegraphics[width=\textwidth]{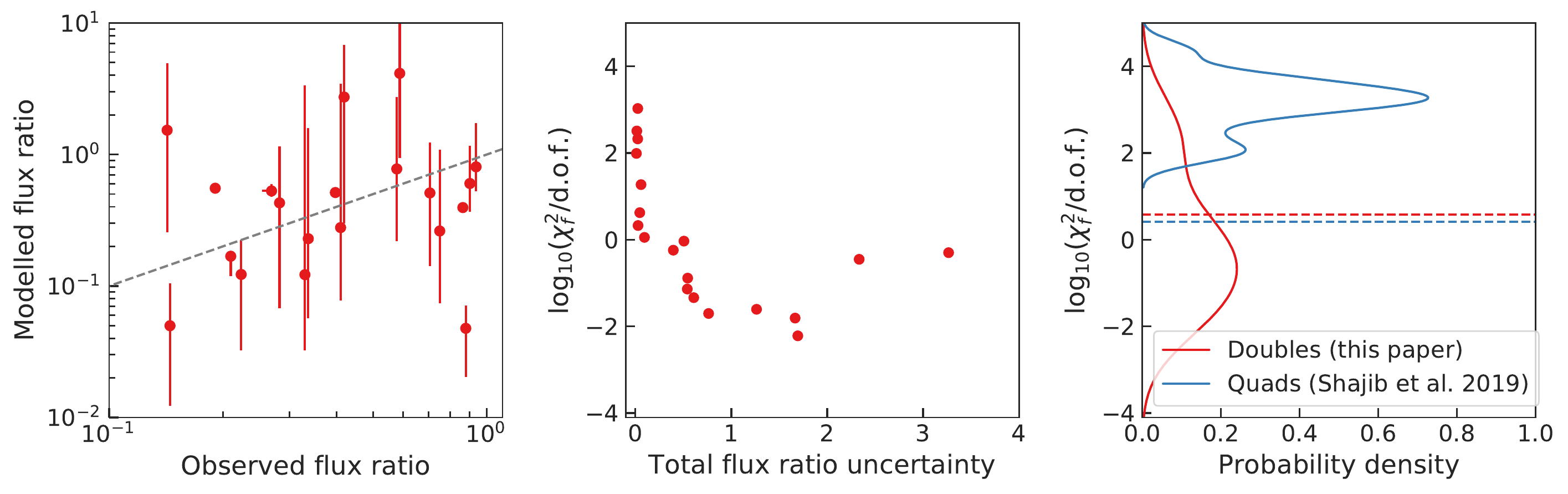}
	\caption{\label{fig:flux_anomaly}
		{\textbf{Left-hand panel:} Comparison of the observed and modelled flux ratios of the doubly imaged systems in this paper. The dashed grey line traces the one-to-one ratio. For most of the lenses, the model uncertainty is much larger than the observed uncertainty, because only two images positions are under-constraining for the model parameters. \textbf{Middle panel:} Correlation between flux ratio departure $\chi^2_{f} \equiv (f_{\rm model} - f_{\rm data})^2/(\sigma_{\rm model}^2 + \sigma_{\rm data}^2)$ from the smooth-model prediction and the total uncertainty $(\sigma_{\rm model}^2 + \sigma_{\rm data}^2)$. The degree of freedom (d.o.f.) for doubly imaged quasars is 1. Expectedly, the lenses with large total uncertainty leads to less departure of the flux ratio from the smooth-model prediction.  \textbf{Right-hand panel:} Distribution of the flux ratio departure from the smooth-model prediction for the doubly imaged quasars (doubles, red) in this study and the quadruply imaged quasars (quads, blue) from \citet{Shajib19}. The d.o.f. for the $\chi^2$ quantity is 3 for quads. The dashed lines show the 95th percentile for the expected distribution for a smooth mass density profile. The distribution for the quads demonstrate a clear departure from the prediction of the smooth model. However, the flux ratios of the doubles is more consistent with the smooth-model prediction due to the large model uncertainty. There is an extended tail in the distribution for doubles toward higher $\chi^2_{f}$, which correspond to the systems with lower model uncertainty and is consistent with the quads' distribution in departing from the smooth-model prediction.}
	}
\end{figure*}

\editref{We compare the observed and model-predicted flux ratios between the quasar images in Figure \ref{fig:flux_anomaly}. The observed flux ratio can depart from the model-predicted one due to microlensing by foreground stars or mililensing by dark matter substructure, due to dust extinction, or due to the arrival time-delay coupled with intrinsic variability [see \citet{Schneider06} for a detailed description]. We quantify the departure of the observed flux ratio from the model-predicted one with a $\chi^2$ quantity
\begin{equation}
	\chi^2_f \equiv \frac{(f_{\rm model} - f_{\rm data})^2}{\sigma_{f, {\rm model}}^2 + \sigma_{f, {\rm data}}^2},	
\end{equation}
where $f$ is the flux ratio $f \equiv F_{\rm dimmer}/F_{\rm brighter}$ with $F$ being the flux of the lensed image, and $\sigma_{f}$ is the uncertainty on the flux ratio. Due to the large model uncertainty in the majority of the lenses, the $\chi^2_{ f}$ quantity is mostly consistent with the smooth-model prediction (Figure \ref{fig:flux_anomaly}). In comparison, the quadruply imaged quasar systems (quads) from \citet{Shajib19} show clear departure from the smooth-model prediction, as quads have tighter constraints on the lens model. However, the doubles that have tight constraint on the lens model do show clear departure from the smooth-model prediction creating an extended tail toward larger values in the $\chi^2_f$ distribution. Thus, given similar model uncertainty, doubles and quads have comparable departure in the observed flux ratio from the smooth-model prediction, which agrees with the prediction by \citet{Schechter02}.
}

\begin{figure}
	\includegraphics[width=\columnwidth]{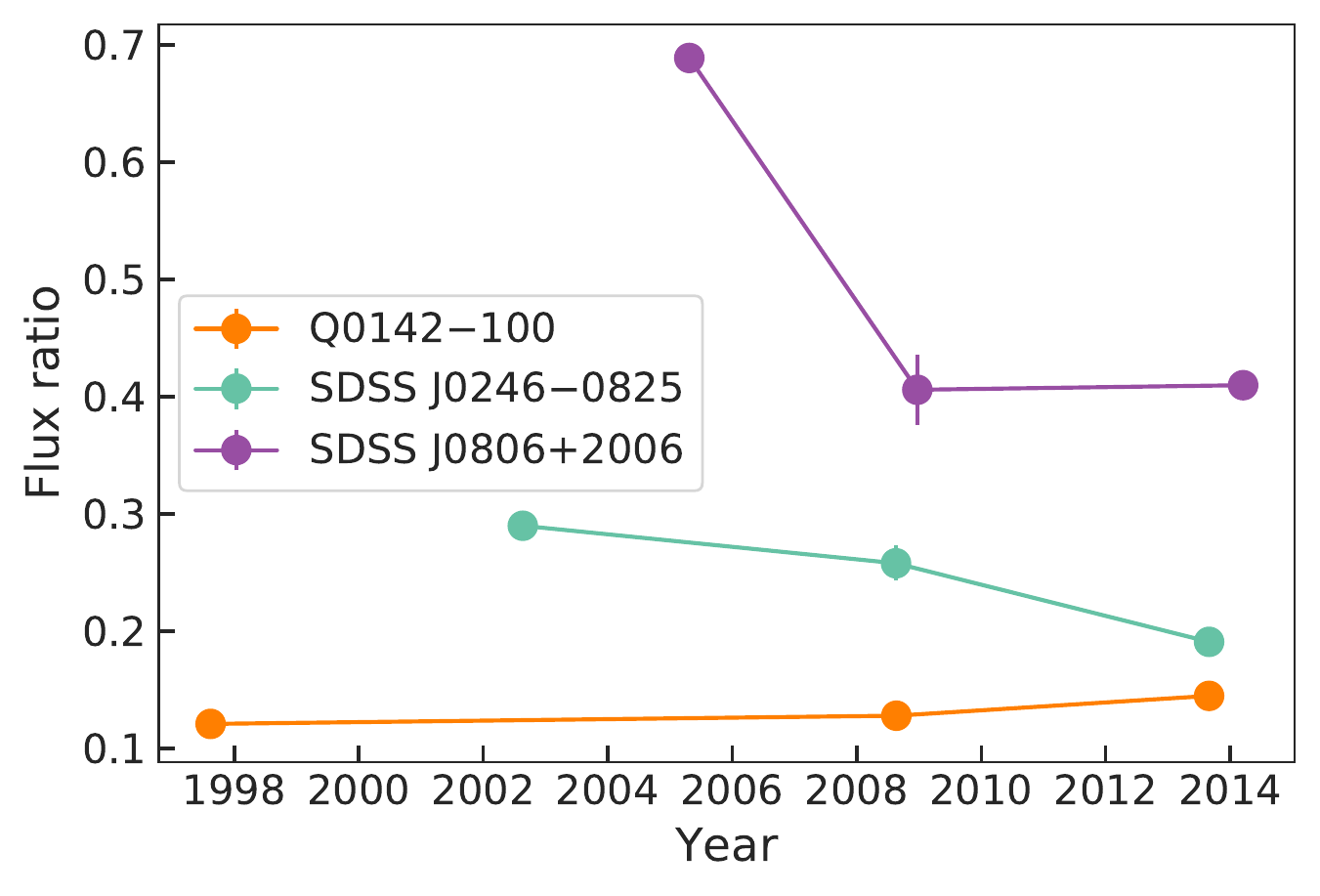}
	\caption{\label{fig:flux_ratio_history}
	Comparison of our measured flux ratios of three lens system with previous observations. Our observed values are the right-most points for each lens. The middle values are from $K$-band observation with Gemini North 8-m telescope \citep{Fadely11}. The left-most observation for Q0142$-$100 is from the \textit{Hubble Space Telescope} observation in the F160W filter \citep{Lehar00}. The left-most observations for SDSS J0246$-$0825 and SDSS J0806$+$2006 are from $K^\prime$-band observation with NIRC on the Keck telescope \citep{Inada05, Inada06}. Our observed values are largely consistent with those from \citet{Fadely11}, whereas the observed variations can be attributed to the movement of the foreground stars causing a variation in the microlensing magnification.
	}
\end{figure}

\editref{We compare our observed flux ratios in the three systems -- Q0142$-$100, SDSS J0246$-$0825, and SDSS J0806$+$2006 -- with previously measured values in $K$-band, $K^\prime$-band, or $H$-band (F160W filter of the \textit{Hubble Space Telescope}) in Figure \ref{fig:flux_ratio_history}. Our observed values are largely consistent with the ones from the most recent past observation of \citet{Fadely11}. The variation in the flux ratio over 5--10 years baseline can be explained by the change in the microlensing magnification pattern due to the movement of the foreground stars in the deflector galaxy.
}

\editrefs{Our relative astrometry between the quasar images are discrepant by $\sim$0.02 arcsec from previous AO-assisted observations of SDSS J0806$+$2006 \citep{Sluse08} and SDSS J1001$+$5027 \citep{Rusu16}. Such a discrepancy can potentially arise from incorrect centering in our sub-optimal PSF. This discrepancy level is negligible for planning future observations, however caution should be taken when using our reported astrometry in studies sensitive to the astrometric accuracy.}

\begin{figure}
	\includegraphics[width=\columnwidth]{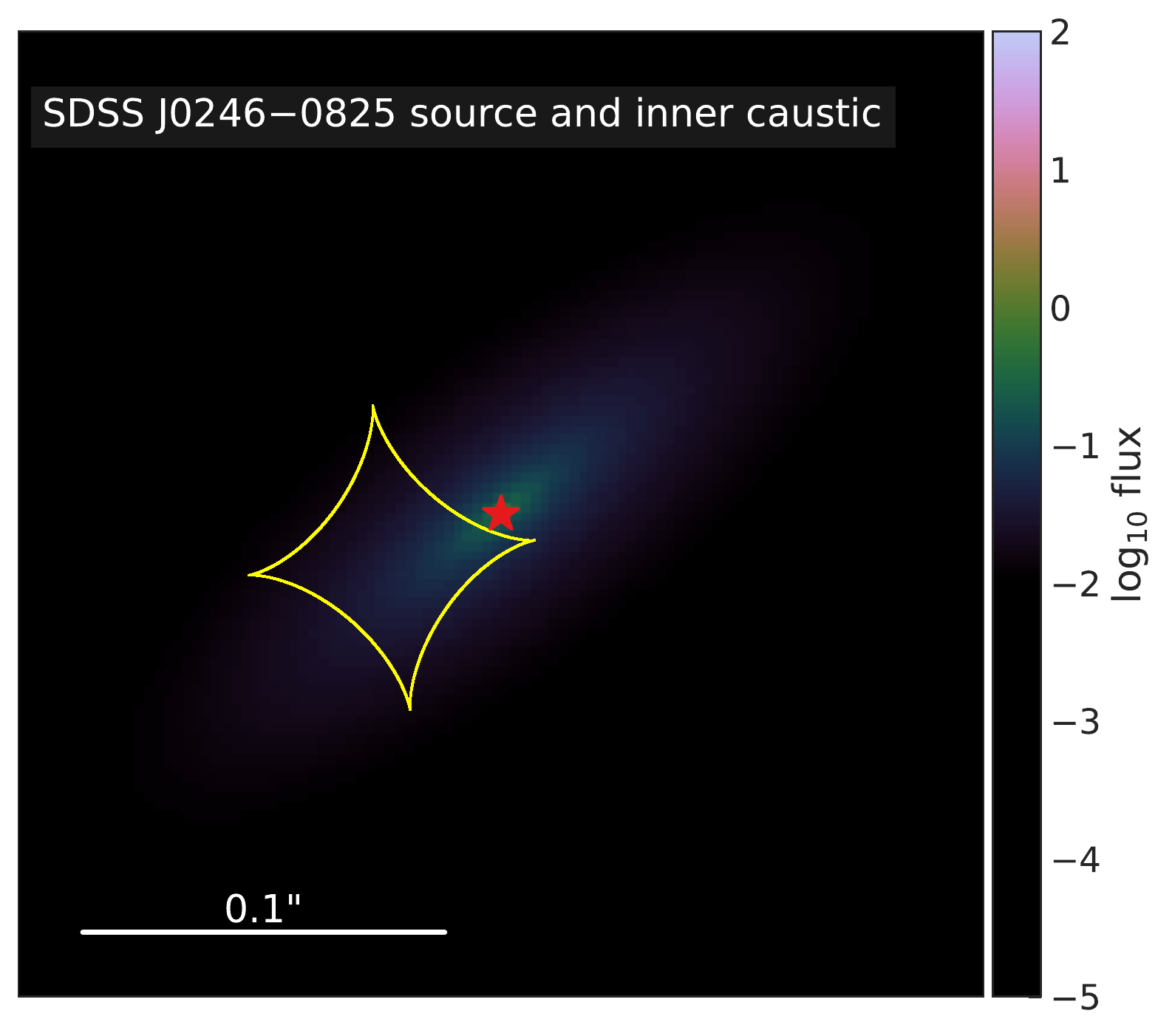}
	\caption{
	\label{fig:0246_source}
	Reconstructed flux distribution of the quasar host galaxy in the SDSS J0246$-$0825 system. The red star marks the position of the central quasar. The yellow line traces the inner caustic.}
\end{figure}

\editref{Lens model of the SDSS J0246$-$0825 system from \citet{Inada05} constrained only by the image positions and the flux ratio
	 suggested that there might be a small faint lensing object near the
	 primary lensing galaxy.  This model closely traced the lensed arc, but no
	 attempt was made to model its intensity. In contrast, no second lensing galaxy is
	 seen in the data presented in this paper, but our approach models the ring intensity
	 and as a result, reconstructs the host galaxy.  As illustrated in Figure \ref{fig:0246_source},
	 the quasar is quite close to the inner diamond-shaped caustic, and small changes to
	 our model would result in a four image system rather than two, with
	 two new images appearing at the brightest spot on the lensed arc.}

\editref{For time-delay cosmography, an ideal double would require (i) prominent lensed arcs to provide tight model constraint on the mass density profile, and (ii) a long time delay to minimize the fractional uncertainty in the time-delay measurement. Although a number of doubles in our sample have noticeable lensed arcs, none of these systems have ideally long ($\sim$100 days) time delays. Out of the systems with noticeable lensed arcs, WISE 2329--1258 has the longest predicted time-delay with $\sim$25 days. This is comparable to the time delay of PG 1115$+$080 with 8.4 per cent uncertainty on a single time-delay measurement, which has been analyzed to measure the Hubble constant \citep{Bonvin18, Chen18}. Since the modelling uncertainty for one double is also $\sim$8.5 per cent \citep{Birrer19, Wong20}, this system would lead to  a $\sim$12--13 per cent Hubble constant measurement assuming a 3--5 per cent uncertainty coming from the external convergence estimate. However, the large modelling uncertainty of the double SDSS 1206$+$4332 largely stemmed from a number of nearby perturber galaxies, the modelling uncertainty for other doubles with less crowded nearby environment would potentially be tighter than $\sim$8.5 per cent.
}

\section{Data availability}
The NIRC2 data used in this paper are publicly available from the Keck Observatory Archive. The lens modelling software \textsc{lenstronomy} used in this paper is an open-source software publicly available on Github.

\section*{Acknowledgments}

\editrefs{We thank the anonymous referee for very useful comments that helped us improve this work.} AJS thanks Abhimat K. Gautam for helpful discussion and Dominique Sluse for useful comments on the manuscript. AJS, TT\editast{, and CDF were} supported by National Aeronautics and Space Administration (NASA) through the Space Telescope Science Institute (STScI) grand HST-GO-15320. AJS was additionally supported by the Dissertation Year Fellowship from the University of California, Los Angeles (UCLA) Graduate Division. This research was supported by the U.S. Department of Energy (DOE) Office of Science Distinguished Scientist Fellow Program. \editas{TT acknowledges support by the National Science Foundation through grant AST-1906976\editast{,} and by the Packard Foundation through a Packard Research Fellowship.} \editast{CDF acknowledges support by the National Science Foundation through grant AST-1907396.}

\editas{AJS acknowledges the hospitality of the Aspen Center of Physics (ACP), where part of this research was completed. ACP is supported by National Science Foundation grant PHY-1607611.}

Some of the data presented herein were obtained at the W.M. Keck Observatory, which is operated as a scientific partnership among the California Institute of Technology, the University of California and the National Aeronautics and Space Administration. The Observatory was made possible by the generous financial support of the W.M. Keck Foundation. The authors wish to recognize and acknowledge the very significant cultural role and reverence that the summit of Mauna Kea has always had within the indigenous Hawaiian community. We are most fortunate to have the opportunity to conduct observations from this mountain, and we respectfully say mahalo.

This research made use of \textsc{lenstronomy} \citep{Birrer15, Birrer18}, \textsc{cosmohammer} \citep{Akeret13}, \textsc{fastell} \citep{Barkana99}, \textsc{numpy} \citep{Oliphant15}, \textsc{scipy} \citep{Jones01}, \textsc{astropy} \citep{AstropyCollaboration13, AstropyCollaboration18}, \textsc{jupyter} \citep{Kluyver16}, \textsc{matplotlib} \citep{Hunter07}, \textsc{pandas} \citep{McKinney10}, and \textsc{seaborn} \citep{Waskom14}.



\bibliographystyle{mnras}
\bibliography{../../ajshajib_old}


\appendix

%
%
%

\section{Outcomes of imaging confirmation campaigns} \label{app:iamgin_campaign}
Tables~\ref{tab:2016} and \ref{tab:2017_2018} list the systems that were observed with NIRC2 on 2016 \editas{September,} 2017 October, and 2018 January, with their coordinates, outcomes, and parent surveys. Contaminants\editas{ -- }labeled `cont.'\editas{ -- }are objects that are revealed as non-lenses \editast{(e.g., star and quasar, or point source and galaxy)} either through our high-resolution imaging or spectroscopy. Nearly identical quasar (NIQ) pairs are putative lenses where no sign of the deflector is found in imaging data \citep{Schechter17, Agnello18b}. They could be veritable lenses with faint deflectors, or close pairs of quasars at the same redshift, \editas{with} same lines and monotonic flux-ratios with wavelength.

\begin{table}
  \caption{\label{tab:2016}
  Summary of \editas{2016 September} imaging observations. Object coordinates are given in the first two columns, followed by imaging classification and parent survey. The lens HE 0013$-$2542 has been selected by P.L.~Schechter in the VST-ATLAS survey, as a possible quasar pair or lens, starting from a sample of Hamburg-ESO \editas{(HE)} quasars. Parent Survey shorthand\editas{ -- A: VST-ATLAS, D: DES; P: Pan-STARRS1, S: SDSS.}}
\begin{tabular}{lrrll}
\hline
Name & RA & Dec & outcome & Parent  \\
 & (deg) & (deg) &  &  Survey \\
 \hline 
J0001+1411 & 0.31665 & 14.18974 & cont. & S \\ 
J0005+2031 & 1.49748 & 20.52355 & cont. & S \\
J0013$-$2542 & 3.93292 & -25.43806 & Lens & HE \\ 
J0024+0032 & 6.1838018 & 0.53931 & cont. & S \\ 
J0037+0111 & 9.33269 & 1.18742 & cont. & S \\ 
J0048+2505 & 12.14571 & 25.08965 & cont. & S \\ 
J0252+3420 & 43.07300 & 34.33824 & inconcl. & S \\ 
J0252$-$0855 & 43.08799 & -8.92101 & \editas{inconcl.} & S \\ 
J0340+0057 & 55.19833 & 0.95997 & cont. & S \\ 
J0502+1310 & 75.61556 & 13.18222 & cont. & S \\ 
J1700+0058 & 255.10005 & 0.97087 & cont. & S \\ 
J1704+1817 & 256.13547 & 256.13547 & cont. & S \\ 
J1738+3222 & 264.70178 & 32.37678 & cont. & S \\ 
J1810+6344 & 272.51841 & 63.74072 & cont. & S \\ 
J2036$-$1801 & 309.21955 & -18.02927 & cont. & S \\  
J2044+0314 & 311.20357 & 3.24864 & cont. & S \\ 
J2045$-$0101 & 311.40236 & -1.03000 & cont. & S \\ 
J2055$-$0515 & 313.87530 & -5.25045 & cont. & S \\  
J2103+1100 & 315.84197 & 11.00532 & cont. & S \\ 
J2111$-$0012 & 317.78773 & -0.21647 & cont. & S \\ 
J2121$-$0005 & 320.37559 & -0.09087 & cont. & S \\ 
J2123$-$0050 & 320.87278 & -0.84804 & cont. & S \\ 
J2146+0009 & 326.55547 & 0.15857 & cont. & S \\ 
J2158+1526 & 329.67363 & 15.43747 & cont. & S \\ 
J2209+0045 & 332.27882 & 0.76218 & cont. & S \\ 
J2238+2718 & 339.53716 & 27.31368 & cont. & S \\
J2246+3118 & 341.69171 & 31.30472 & cont. & S \\
J2257+2349 & 344.35586 & 23.82510 & Lens & S \\
J2329$-$1258 & 352.49125 & -12.98306 & Lens & A, P \\
J2350$-$0749 & 357.51073 & -7.82589 & cont. & S \\
J2353$-$0539 & 358.46255 & -5.66552 & cont. & S \\
J2358$-$0136 & 359.58584 & -1.60291 & cont. & S \\

 \hline 
  \end{tabular}
\end{table}

\begin{table}
 \caption{\label{tab:2017_2018}
 Summary of 2017 October and 2018 January imaging observations. Object coordinates are given in the first two columns, followed by imaging classification and parent survey. \editas{Short-hands for parent surveys are the same as in Table \ref{tab:2016}.}
 } 
 \begin{tabular}{lrrll}
 \hline
Name & RA & Dec & outcome & Parent  \\
 & (deg) & (deg) &  &  Survey \\

 \hline 
 
J0116+1446 & 19.06358 & 14.77958 & cont. & P \\ 
J0146$-$1133 & 26.63708 & -11.56083 & Lens & A, D, P \\
J0235$-$2433 & 38.86426 & -24.55368 & Lens & A, D, P  \\ 
J0259$-$2338 & 44.88965 & -23.63383 & Lens & A, D, P  \\ %
{J2003$-$2111} & 300.75063 & -21.18501 & \editas{inconcl.} & P \\  
J2029$-$0706 & 307.33973 & -7.10646 & cont. & P \\
J2041+0722 & 310.39218 & 7.37083 & cont. & P \\
J2057+0217 & 314.46716 & 2.29670 & cont. & S, P \\
J2155+1903 & 328.75684 & 19.05074 & cont. & P \\
J2213$-$2652 & 333.41012 & -26.87419 & Lens & A \\
{J2303+3453} & 345.91142 & 34.89518 & \editas{inconcl.} & P \\ 
J2305+3714 & 346.48239 & 37.23899 & Lens & P \\

%
 \hline 
 
J0140+4107 & 25.20420 & 41.13330 & Lens & P \\ 
J0245$-$0556 & 41.35651 & -5.95015 & Lens & D \\ 
J0407$-$1931 & 61.97413 & -19.52254 & cont. & D \\ 
J0417+3325 & 64.49683 & 33.41700 & Lens & P \\ 
J0723+4739 & 110.93660 & 47.65260 & cont. & P \\
J0740+2926 & 115.05603 & 29.44677 & cont. & P \\
J0812+3349 & 123.22844 & 33.83062 & cont. & P \\
J0840+3550 & 130.13842 & 35.83334 & Lens & P \\
J0949+4208 & 147.47830 & 42.13381 & Lens & P \\
LBQS~1009$-$0252 & 153.06625 & -3.11750 & known & A \\
J1112$-$0335 & 168.18096 & -3.58592 & NIQ & A \\
J1128+2402 & 172.07705 & 24.03820 & Lens & S \\
J1132$-$0730 & 173.03091 & -7.51178 & NIQ & A \\

%
%
%
 \hline
  \end{tabular}
\end{table}


\bsp	
\label{lastpage}
\end{document}